# Hole Spin Pumping and Re-pumping in a p-type δ-doped InAs Quantum Dot

Konstantinos G. Lagoudakis[†], Kevin A. Fischer, Tomas Sarmiento, Kai Mueller
& Jelena Vučković

*E. L. Ginzton Laboratory, Stanford University, Stanford, California 94305, USA*

We have grown high quality p-type δ-doped InAs quantum dots and have demonstrated coherent spin pumping and repumping of a hole spin in a positively charged quantum dot by means of a single-laser driving scheme under a high magnetic field in the Voigt configuration. Modeling of our system shows excellent qualitative agreement with the experimental findings and further explores the performance of the single-laser scheme for spin pumping and re-pumping.

Quantum information processing relies on robust quantum bits that feature long coherence times and immunity to the surrounding environment. The electron spin in charged quantum dots (QDs) has been successfully used for quantum information applications[1] but is limited by its short coherence time due to strong hyperfine interactions with the surrounding nuclear spin bath[2]. The *p*-symmetry of the heavy-hole Bloch wavefunction significantly reduces the contact hyperfine interaction with the surrounding nuclei[3,4] making the hole spin in positively charged quantum dots a very attractive and robust candidate for the implementation of qubits with long coherence times[5,6,7]. Previous studies of hole spin initialization[8,9,10,11,12] and coherent control[6,7,13,14] predominately relied on tuneable p-i-n or Schottky diode structures for quantum dot charge control. In these structures charging occurs either by voltage-controlled tunnelling of charge carriers into the QD from a close by reservoir[15] or by tunnel-ionisation of photogenerated excitons[14]. Although these methods reliably charge the QDs, they suffer from noise that is induced by the electric field fluctuations[16,17], can be subject to spin decoherence from the nearby reservoir and they require a multi step process for charging of the QDs[9]. δ-doped samples on the other hand do not require the same level of complex nanofabrication or suffer from the aforementioned drawbacks. However, δ-doped samples have additional challenges in their growth, most notably the introduction of impurities, which is detrimental to the optical properties of the dots. Thus, successful growth of high quality δ-doped samples is highly favorable for use in quantum information processing.

Here, we have grown high quality p-type δ-doped InAs quantum dots, and we demonstrate coherent spin pumping and repumping of a hole spin in a positively charged quantum dot utilizing a single-laser scheme. We accurately model the observed phenomenology, providing insight into the extent of the parameter space for which the single-laser scheme is valid.

The QD sample used in this work was grown by molecular beam epitaxy on a GaAs substrate and consisted of a single layer of InAs QDs emitting around 890 nm and capped by a 100 nm GaAs layer. Positive charging of the dots was realized by a Beryllium δ-doped layer inserted 10 nm below the QD layer. A high doping density of $\sim 10^{11}/cm^2$ was used to ensure high charging probability, with more than half of the QDs being in a charged state. Despite this high doping density, the QDs exhibit relatively narrow linewidths ranging from 10 to 40μeV. Charging of the dots was probed under the application of a magnetic field in the Voigt configuration (magnetic field perpendicular to growth axis), resulting in the quadruplet splitting of the excitonic transitions because of the Zeeman interaction. Application of the magnetic field was done with our recently developed magnetospectroscopy setup[18] and spectrally resolved studies were enabled by a custom made dual monochromer of 1.75m in overall length, boasting a ~10μeV resolution. Our setup's high suppression ratio allowed detection of excitonic transitions in close proximity to resonant lasers (as close as 0.05nm). A statistical analysis on 50 charged dots from our sample showed that the g factors of electrons and holes are distributed with $g_e$=0.32±0.03 and $g_h$=0.15±0.07, where the hole g factors are significantly more inhomogeneous, consistent with previous literature results[19]. Thanks to this large inhomogeneity, several dots with $g_e \approx g_h$ can be found, which is critical to the single-laser experiment presented here. A typical quantum dot with $g_e \approx g_h$ was initially investigated at zero magnetic field under above band excitation ($\lambda_{ab} \approx 780$nm) and in saturation, in order to extract the spectral linewidth, here 22μeV (0.014nm), and the fine structure splitting (FSS) of 1.8μeV (0.001nm)) as shown in Fig. 1(a). Such low FSS indicates very low stress/strain in the vicinity of the QD - beneficial for many quantum information processing applications[20]. Additionally, the luminescence intensity (Fig. 1(b)) showed a linear dependence for low excitation power followed by saturation for powers beyond 4μW, indicating that the transition under investigation is either from a neutral or singly charged exciton. To further probe the existence of an excess hole and therefore the positive charging of the investigated quantum dot, a spectroscopic investigation was carried out under high magnetic field in the Voigt configuration.

For a positively charged dot, the ground state is occupied by a heavy hole with ±3/2 spin and the excited state is occupied by a spinless hole pair and an electron with overall spin ±1/2. The linear in B Zeeman splitting $\delta_h/\delta_e$ of the ground/excited state for a magnetic field applied in the Voigt configuration is proportional to the g factor of the hole/electron $\delta_{h/e}=\mu_B g_{h/e} B$ where $\mu_B$ is the Bohr magneton and B is the applied magnetic field. Although the diamagnetic shift (an additional quadratic term in B) is always present, it can be easily removed to reveal the pure Zeeman splittings. These splittings transform the charged QD into a double Λ system with a characteristic quadruple spectral line signature, while the selection rules governing these four levels impose

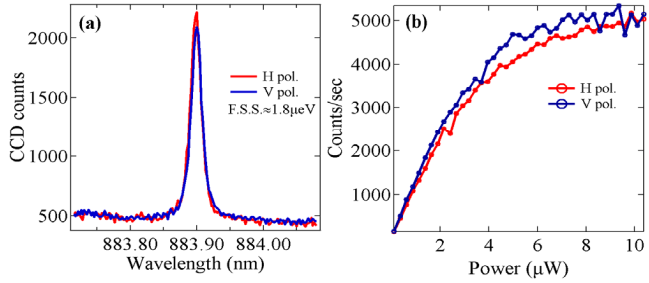

FIG. 1: (a) Quantum dot spectra in horizontal and vertical polarizations at zero magnetic field. Fitting of the peaks reveals a linewidth of the order of ~22μeV and a very small fine structure splitting of the order of ~1.8μeV. (b) Intensity dependence of the H and V transitions as a function of the pumping power for above band excitation, showing a linear low-power region followed by intensity saturation for higher powers.

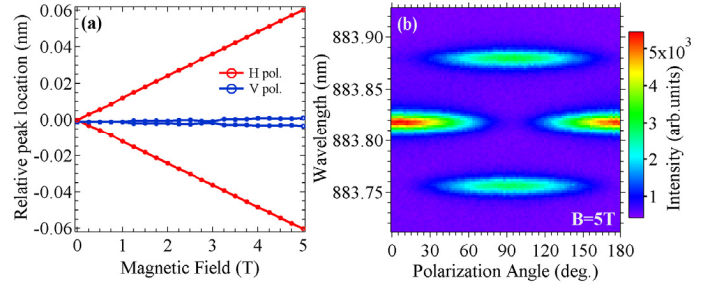

FIG. 2: (a) Spectroscopic study of the excitonic transitions as a function of the magnetic field intensity in the Voigt configuration. Multi-peak fitting reveals that the excitonic lines develop into the characteristic quadruplet structure (because of the Zeeman interaction) with the two inner transitions almost degenerate, indicating that the g factors for electron and holes are approximately equal. (b) Complete polarization analysis for our maximum reachable magnetic field ($B_{max}$=5T) under above-band excitation. The inner transition corresponds to a pair of almost degenerate emission lines giving twice the intensity of each of the two outermost transitions.

horizontal/vertical polarization for the two outer/inner transitions[21].

As we gradually increased the magnetic field (in the Voigt configuration), the excitonic resonance of the QD split into a quadruple peak indicating that the dot is charged. Linear regression multi-peak fitting allowed us to track the relative positions of these peaks as a function of the field, and after elimination of the diamagnetic shift, we show the pure Zeeman splitting in Fig. 2(a). The diamagnetic shift factor was here found to be 5.07±0.03μeV/T². From the energies of the four peaks at maximum field we also extracted the |g| factors, here $g_{e/h}$ ≈ 0.34±0.02. A complete polarization analysis for our maximum reachable magnetic field $B_{max}$=5T in Fig. 2(b) shows the spectral/polarization structure of the quantum dot with the inner transition amplitude being twice as high as the two outer transitions, indicative of the presence of two almost degenerate peaks. A close inspection of Figs. 1(a) and 2(b) reveals a mismatch of the wavelength of the central peak that is here due to the Diamagnetic shift.

We illustrate this double Λ system in Fig. 3(a) with the wavy downward arrows depicting the two outer horizontally polarized transitions and the two upward straight arrows depicting the two inner vertically polarized transitions. Application of the vertically polarized 'pump' laser in Fig. 3(a) brings the system from state $|\Downarrow\rangle$ to state $|\Uparrow\Downarrow\uparrow\rangle$ while spontaneous emission moves the system back to one of the two ground states $|\Downarrow\rangle$ or $|\Uparrow\rangle$ with equal probabilities. If the system decays back to $|\Downarrow\rangle$, then the pump laser will re-excite the system to $|\Uparrow\Downarrow\uparrow\rangle$ whereas if it decays to $|\Uparrow\rangle$, a horizontally polarized photon is detected and the system becomes initialized to the $|\Uparrow\rangle$ state. In this scenario, after the first photon is detected, the system reaches its steady state and no more detection events are recorded. If a second 'repump' laser is applied, then the system is brought from $|\Uparrow\rangle$ to $|\Uparrow\Downarrow\downarrow\rangle$ and the same decay scenarios will occur. If both pump and repump lasers are kept on, they cycle population through all four states resulting in the detection of a continuous stream of photons.

For a charged quantum dot with $g_e\neq g_h$, the two inner vertical transitions are at different energies ($\delta_h\neq\delta_e$ in Fig. 3(a)) and therefore two independent lasers are required for pumping and repumping[22]. However, for quantum dots with $g_e\approx g_h$ like the one studied here, the two inner transitions are degenerate ($\delta_h\approx\delta_e$) and therefore a single-laser should be sufficient for the simultaneous spin pumping and repumping processes.

To probe the expected effect, a tunable single-mode Ti: Sapphire laser was brought close to resonance with the two doubly degenerate inner transitions and the wavelength was finely scanned over ~0.04nm range. Meanwhile we recorded both the exact wavelength of the resonant laser using a precision wavemeter as well as the detected single-photon counts from the high energy outer transition (Fig. 3(a)) using a single-photon counting module at the output of the dual monochromator. The use of the monochromator was here required to spectrally filter photons coming from this transition while suppressing the nearly resonant reflected laser. We show the detected number of photon counts as a function of the exact laser wavelength in Fig. 3(b) for 2μW incident power. A clear resonance can be seen with a 16μeV (0.01nm) FWHM. A power dependence study revealed a strong effect on the width of the detected resonance. When increasing the power from 0.1μW to 10μW, the width of the resonance shows a nearly linear behavior with an almost threefold increase in the resonance width from 11.5μeV to 36μeV (0.007nm to 0.023nm). This behavior is shown in Fig. 3(c) and is attributed to power broadening[23]. For higher powers, the error bars in Fig. 3(c) are more pronounced because of the increased photon leakage of the laser in the detection channel. A closer look to the peak height of the resonance shows a strong increase for low excitation power followed by a saturation plateau above 1μW of incident power as shown in Fig. 3(d).

Intuitively, as long as the difference of the energy splittings $\delta_h$ and $\delta_e$ is within a power-broadened natural linewidth of these states, some spectral overlap between the two vertically polarized transitions exists and a single laser should be sufficient in order to pump and repump the hole spin. To confirm this assumption and better understand the extent to which it holds we performed precise modeling of a four level system for an extended range of values in the parameter space using the quantum optics toolkit (QuTip)[24] in Python.

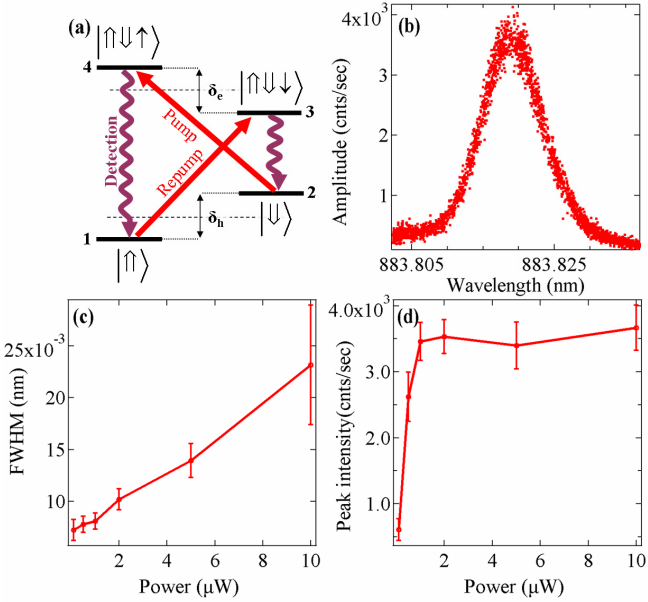

FIG. 3: (a) Level structure of the positively charged QD at high magnetic field. For a dot with $g_e \approx g_h$ the Zeeman splitting for the ground and excited states is almost equal which results in two equi-energetic inner (vertically polarized) transitions, here denoted by the two red arrows labelled "Pump" and "Repump". Detection events are counted only from the highest energy (horizontally polarized) transition labelled "Detection" with high resolution spectral filtering. (b) Rate of detection events as a function of the wavelength of the pump-repump resonant laser for 2μW excitation power. (c) Spectral width and (d) peak intensity of the resonance as a function of the driving laser power.

The Hamiltonian representing our 4 level system driven by the pump-repump laser is $\mathcal{H} = \mathcal{H}_o + \mathcal{H}_d$ where,

$$\mathcal{H}_o = -\frac{\delta_h}{2}\Pi_1 + \frac{\delta_h}{2}\Pi_2 + \left(\omega_o - \frac{\delta_e}{2}\right)\Pi_3 + \left(\omega_o + \frac{\delta_e}{2}\right)\Pi_4 \quad (1)$$

is the unperturbed four level system Hamiltonian and the driving term is given by

$$\mathcal{H}_d = \Omega e^{i\omega_l t}(\sigma_{13} + \sigma_{24}) + h.c. \quad (2)$$

The self energy terms are contained in $\mathcal{H}_o$ where $\Pi_i = |\varphi_i\rangle\langle\varphi_i|$ are each level's respective projection operator while $\omega_o$ is the zero-field level splitting. The QD driving component $\mathcal{H}_d$, models the single-laser (with frequency $\omega_l$) spin pump and repump interaction under the rotating wave approximation. The Rabi frequency governing the strength of this interaction is given by $\Omega$ which is proportional to the driving electric field magnitude.

Here we are interested in the spontaneous emission intensity from the highest energy transition, proportional to $\langle\Pi_4\rangle$, and hence we wish to compute the average population levels in the long time limit. Thus, transformation to a rotating frame in order to yield the time-independent Hamiltonian $\tilde{\mathcal{H}}$ greatly simplifies this calculation. We simply look for the steady state of the density matrix $\tilde{\rho}$ in the rotating frame, i.e. solve

$$\frac{d\tilde{\rho}(t\to\infty)}{dt} = \left\{-i\left[\tilde{\mathcal{H}},\tilde{\rho}(t\to\infty)\right] + \sum_j \mathcal{L}(c_j)\right\} \to 0 \quad (3)$$

where $\mathcal{L}(c_j)$ are the Lindblad superoperators for each collapse operator $c_j$ of the system. Here, we have only included longitudinal dipole decay terms (with rates commonly used in similar systems[25,26]). The calculated steady state density matrix $\tilde{\rho}(t\to\infty)$ directly gives the relative photon count rate that we expect to measure.

Fig. 4(a) shows the calculated photon count rate from the pump-repump process under single-laser excitation for a wide range of hole g factors while keeping $g_e$=0.34, as a function of the laser wavelength. For the zero detuning case between the two crossed transitions ($g_e=g_h$) the intensity of the resonance is at its maximum and as $g_h$ grows increasingly different than $g_e$ the resonance amplitude is quickly decreased as the pump and repump cycling of the system becomes more inefficient.

Although for the fully resonant case where $g_e=g_h$ one would intuitively expect a single peak for the pump-repump resonance, it is harder to predict the shape for larger detunings. Fig. 4(b) addresses this question; by normalizing the resonance profile to its peak intensity for each value of $g_h$ from Fig. 4(a), the shape can be easily resolved even for very low intensities. Within the range of $g_h$ that was chosen here, significant broadening of the resonance can be identified and counter-intuitively it always remains singly peaked. Indeed, at first glance one could be tempted to fit this resonance using a simple sum of two displaced Lorentzians but the mechanism of simultaneous spin pumping and repumping dictates that the form of the observed resonance is a result of a multiplication of the two individual displaced Lorentzians.

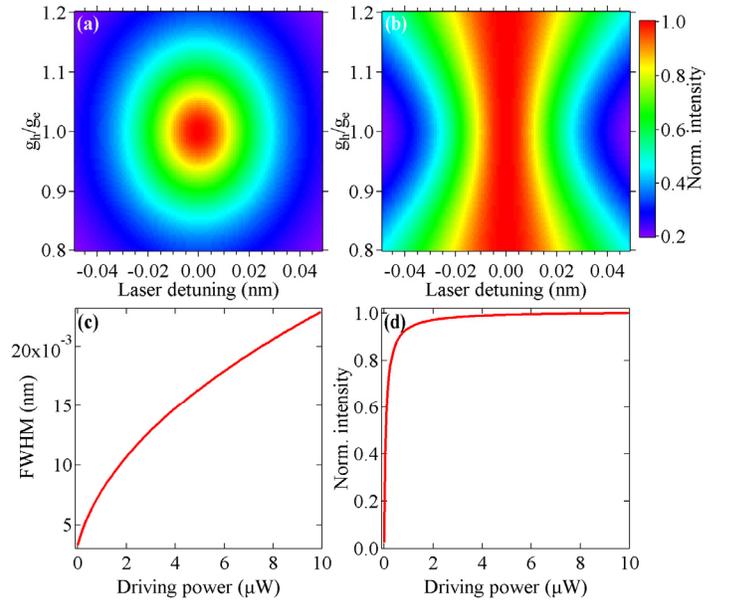

FIG. 4: (a) Simulated pump-repump resonance intensity for a wide range of $g_h$ values relative to $g_e$ =0.34 as a function of laser detuning (in our single laser scheme). (b) Same as in (a) but with each resonance profile normalized to its peak value, showing the broadening of the observed resonance for increasing energy difference between the two inner transitions (c) Power broadening and (d) intensity saturation for the parameters that reproduce the experimentally observed phenomenology (in (a),(b) $\Omega/2\pi$=1.0GHz, $\gamma/2\pi$=0.25GHz and $g_h=g_e$)

Using our model, we also simulated the power broadening and peak intensity behaviors for $g_e=g_h$. Fig. 4(c) shows the power broadening behavior, here evolving quasi-linearly with the driving power. The simulated resonance peak count rate in Fig. 4(d) shows a fast increase for low driving power followed by saturation, in excellent agreement with the experiment. At the highest experimental pump power, the emission from the quantum dot is so strongly power broadened that the linewidth is nearly independent of $\gamma$. As a result, we can map the model's driving strength $\Omega$ to the experimental pump power. Using this correspondence we qualitatively fit the saturation curve with the quantum dot's natural linewidth as a free parameter.

Additional modeling that included the effect of a finite spin-state lifetime revealed that the system's strongest dephasing processes dominate the saturation curve. As a result, the T1 time plays little role in determining the shape of the saturation curve unless it exceeds $1/\gamma$. Given that T1 times are typically on the order of μsec for self-assembled InAs quantum dots, the spin lifetime should play effectively no role in the shape of the saturation curve.

In summary we have successfully grown high quality p-type δ-doped quantum dots and have demonstrated coherent spin pumping and repumping using a single-laser scheme. Using numerical modeling of our system we have reproduced the observed phenomenology and we have investigated the extent of the parameter space to which a single-laser scheme is sufficient for spin pumping and repumping. Follow up experiments are currently in progress towards the demonstration of complete quantum control of hole spins in this type of quantum dots both in the bulk as well as in nanostructured devices.

The authors acknowledge financial support from the Air Force Office of Scientific Research, MURI Center for Multifunctional light-matter interfaces based on atoms and solids and support from the Army Research Office (grant number W911NF1310309). K.G.L. acknowledges support from the Swiss National Science Foundation. K.A.F. acknowledges support from the Lu Stanford Graduate Fellowship. K. M acknowledges support from the Alexander von Humboldt Foundation.

†lagous@stanford.edu